# Bäcklund-Transformation-Related Recursion Operators: Application to the Self-Dual Yang-Mills Equation

C. J. Papachristou * ,  B. Kent Harrison **

*Department of Physical Sciences, Naval Academy of Greece, Piraeus 18539, Greece
E-mail:  papachristou@snd.edu.gr

**Department of Physics and Astronomy, Brigham Young University, Provo, UT 84602, USA
E-mail:  bkentharrison@comcast.net , bkh@byu.edu

**Abstract.** By using the self-dual Yang-Mills (SDYM) equation as an example, we study a method for relating symmetries and recursion operators of two partial differential equations connected to each other by a non-auto-Bäcklund transformation. We prove the Lie-algebra isomorphism between the symmetries of the SDYM equation and those of the potential SDYM (PSDYM) equation, and we describe the construction of the recursion operators for these two systems. Using certain known aspects of the PSDYM symmetry algebra, we draw conclusions regarding the Lie algebraic structure of the "potential symmetries" of the SDYM equation.

## I. Introduction

Recursion operators are powerful tools for the study of symmetries of partial differential equations (PDEs). Roughly speaking, a recursion operator is a linear operator which produces a new symmetry characteristic of a PDE whenever it acts on an "old" characteristic (see Appendix). The concept was first introduced by Olver [1, 2] and subsequently used by many authors (see, e.g., [2, 3] and the references therein). An alternative view, based on the concept of a Bäcklund transformation (BT), was developed in a series of papers by the present authors [4-6] who studied the symmetry problem for the self-dual Yang-Mills equation (SDYM). The general idea is that a recursion operator can be viewed as an auto-BT for the "linearization equation" (or symmetry condition) of a (generally nonlinear) PDE. This idea was later further developed and put into formal mathematical perspective by Marvan [7].

It has been known for some time (see, e.g., Section 7.4 of [3] and the references therein) that, when two nonlinear PDEs are connected by a non-auto-BT, symmetries of either PDE may yield symmetries of the other. This can be achieved by using the original BT to construct another non-auto-BT which relates solutions of the linearization equations of the two PDEs. In the particular case of the SDYM equation, the original BT associates this PDE with the "potential SDYM equation" (PSDYM). The symmetries of the latter PDE can then be used to construct the "potential symmetries" of SDYM [5, 8]. We now attempt to go one step further: Can we find a BT which relates *recursion operators* of two PDEs? Given that, as said above, a recursion operator is itself an auto-BT, what we are after is a BT connecting two auto-BTs, each of which produces solutions of a respective linear PDE (symmetry condition). Thus, we are looking for “a transformation of transformations” rather than a transformation of functions.

Our “laboratory” model will again be SDYM, for good reasons. First, it possesses a rich symmetry structure; second, this PDE has been shown to constitute a sort of prototype equation from which several other integrable PDEs are derived by reduction (see, e.g., [9, 10]). By employing a non-auto-BT that connects SDYM with PSDYM, we will show how symmetries and recursion operators of either system can be associated with symmetries and recursion operators, respectively, of the other system. Moreover, we will prove that the symmetry Lie algebras of these two PDEs are isomorphic to each other. This conclusion is more than of academic importance, since it allows us to investigate the symmetry structure of the SDYM problem by studying the relatively easier PSDYM problem. As an example, we will recover the known infinite-dimensional symmetry algebras of SDYM [11-13] from the symmetry structure of PSDYM [8] and show how these algebras are related to potential symmetries.

## II. The Symmetry Problem for the SDYM-PSDYM System

We write the SDYM equation in the form

$$F[J] \equiv D_{\bar{y}}(J^{-1}J_y) + D_{\bar{z}}(J^{-1}J_z) = 0 \qquad (1)$$

We denote by $x^\mu \equiv y, z, \bar{y}, \bar{z}$ $(\mu = 1, \cdots, 4)$ the independent variables, and by $D_y$, $D_z$, etc., the total derivatives with respect to these variables. We will also use the notation $D_y F \equiv F_y$, etc., for any function *F*. We assume that *J* is *SL*(*N*,*C*)-valued (i.e., $\det J = 1$).

We consider the non-auto-BT

$$J^{-1}J_y = X_{\bar{z}} \, , \quad J^{-1}J_z = -X_{\bar{y}} \qquad (2)$$

The integrability condition $(X_{\bar{y}})_{\bar{z}} = (X_{\bar{z}})_{\bar{y}}$ yields the SDYM equation (1). The integrability condition $(J_y)_z = (J_z)_y$, which is equivalent to

$$D_y(J^{-1}J_z) - D_z(J^{-1}J_y) + [J^{-1}J_y \, , \, J^{-1}J_z] = 0 \, ,$$

yields a nonlinear PDE for the “potential” *X* of (1), called the “potential SDYM equation” or PSDYM:

$$G[X] \equiv X_{y\bar{y}} + X_{z\bar{z}} - [X_{\bar{y}} \, , \, X_{\bar{z}}] = 0 \qquad (3)$$

Noting that, according to (2), $(trX)_{\bar{z}} = [tr(\ln J)]_y = [\ln(\det J)]_y$, etc., we see that the condition $\det J = 1$ can be satisfied by requiring that $trX = 0$ [this requirement is compatible with the PSDYM equation (3)]. Hence, *SL*(*N*,*C*) SDYM solutions correspond to *sl*(*N*,*C*) PSDYM solutions.

Let $\delta J=\alpha Q$ and $\delta X=\alpha \Phi$ be an infinitesimal symmetry of system (2) ($\alpha$ is an infinitesimal parameter). This means that $(J+\delta J, X+\delta X)$ is a solution to the system when $(J, X)$ is a solution. This suggests that the integrability conditions $F[J+\delta J]=0$ and $G[X+\delta X]=0$ are satisfied when the integrability conditions $F[J]=0$ and $G[X]=0$ are satisfied; that is, $J+\delta J$ and $X+\delta X$ are solutions of (1) and (3), respectively. The functions $Q$ and $\Phi$ are *symmetry characteristics* for the above PDEs. Geometrically, the symmetries of system (2) are realized as transformations in the jet-like space of the variables $\{x^\mu, J, X\}$ and the various derivatives of $J$ and $X$ with respect to the $x^\mu$. These transformations are generated by vector fields which, without loss of generality, may be considered "vertical", i.e., with vanishing projections on the base space of the $x^\mu$ [2]. We formally represent these vectors by differential operators of the form

$$V = Q\frac{\partial}{\partial J} + \Phi\frac{\partial}{\partial X} \quad (+\, prolongation\ terms) \tag{4}$$

Consider a function $M(J, X)$. Denote by $\Delta M(J, X)$ the Fréchet derivative [2] of $M$ with respect to $V$ (which in this context is locally the same as the Lie derivative). The infinitesimal variation of $M$ in the "direction" of $V$ is then $\delta M=\alpha\Delta M$. The linear operator $\Delta$ is a derivation on the algebra of all $gl(N,C)$-valued functions. The Leibniz rule is written

$$\Delta(M\,N) = (\Delta M)\,N + M\,\Delta N \tag{5}$$

In particular, for the Lie algebra of $sl(N,C)$-valued functions, the Leibniz rule may also be written as

$$\Delta[M,N] = [\Delta M,N] + [M,\Delta N] \tag{6}$$

By definition, the Fréchet derivatives of the fundamental variables $J$ and $X$ are given by

$$\Delta J = Q\,, \quad \Delta X = \Phi \tag{7}$$

We also note that the Fréchet derivative with respect to a *vertical* vector field commutes with all total derivative operators [2]. Finally, for an invertible matrix $M$,

$$\Delta(M^{-1}) = -M^{-1}(\Delta M)\,M^{-1} \tag{8}$$

(For a discussion of the general symmetry problem for matrix-valued PDEs, see [14].)

We introduce the covariant derivative operators

$$\begin{aligned} \hat{A}_y &\equiv D_y + [J^{-1}J_y\,,\ \ ] = D_y + [X_{\bar{z}}\,,\ \ ] \\ \hat{A}_z &\equiv D_z + [J^{-1}J_z\,,\ \ ] = D_z - [X_{\bar{y}}\,,\ \ ] \end{aligned} \tag{9}$$

where the BT (2) has been taken into account. By using (3) and the Jacobi identity, the zero-curvature condition $[\hat{A}_y, \hat{A}_z]=0$ is shown to be satisfied, as expected in view of the fact that the "connections" $J^{-1}J_y$ and $J^{-1}J_z$ are pure gauges. Moreover, the linear operators of (9) are derivations on the Lie algebra of *sl*(*N*,*C*)-valued functions, satisfying a Leibniz rule of the form (6):

$$\begin{aligned} \hat{A}_y\,[M,N] &= [\hat{A}_y M, N] + [M, \hat{A}_y N] \\ \hat{A}_z\,[M,N] &= [\hat{A}_z M, N] + [M, \hat{A}_z N] \end{aligned} \qquad (10)$$

If Eqs. (1)-(3) are satisfied, then so must be their Fréchet derivatives with respect to the symmetry vector field *V* of (4). We now derive the symmetry condition for each of the above three systems. For SDYM (1), the symmetry condition is $\Delta F[J]=0$, or

$$D_{\bar{y}}\,\Delta(J^{-1}J_y) + D_{\bar{z}}\,\Delta(J^{-1}J_z) = 0 \qquad (11)$$

(since the Fréchet derivative $\Delta$ commutes with total derivatives). By using (5), (7), (8) and (9), it can be shown that

$$\Delta(J^{-1}J_y) = \hat{A}_y\,(J^{-1}Q)\,, \quad \Delta(J^{-1}J_z) = \hat{A}_z\,(J^{-1}Q) \qquad (12)$$

The SDYM symmetry condition (11) then becomes

$$(D_{\bar{y}}\hat{A}_y + D_{\bar{z}}\hat{A}_z)(J^{-1}Q) = 0 \qquad (13)$$

The symmetry condition for PSDYM (3) is $\Delta G[X]=0$, or, by using (6), (7) and (9),

$$\hat{A}_y \Phi_{\bar{y}} + \hat{A}_z \Phi_{\bar{z}} \equiv (\hat{A}_y D_{\bar{y}} + \hat{A}_z D_{\bar{z}})\Phi = 0 \qquad (14)$$

We note the operator identity

$$\hat{A}_y D_{\bar{y}} + \hat{A}_z D_{\bar{z}} = D_{\bar{y}}\hat{A}_y + D_{\bar{z}}\hat{A}_z \qquad (15)$$

which is easily verified by letting the right-hand side act on an arbitrary function *M*. Then, (14) is written in the alternate form,

$$(D_{\bar{y}}\hat{A}_y + D_{\bar{z}}\hat{A}_z)\Phi = 0 \qquad (16)$$

Comparing (13) and (16), we observe that the symmetry characteristic $\Phi$ of PSDYM, and the function $J^{-1}Q$, where $Q$ is an SDYM symmetry characteristic, satisfy the same symmetry condition. We thus conclude the following (see also [5]):

- If $Q$ is an SDYM characteristic, then $\Phi = J^{-1}Q$ is a PSDYM characteristic.

Conversely,

- if $\Phi$ is a PSDYM characteristic, then $Q=J\Phi$ is an SDYM characteristic.

Finally, the Fréchet derivative with respect to $V$ also leaves the system of PDEs (2) invariant: $\Delta(J^{-1}J_y)=(\Delta X)_{\bar{z}}$ , $\Delta(J^{-1}J_z)=-(\Delta X)_{\bar{y}}$ . With the aid of (12) and (7) we are thus led to a pair of PDEs,

$$\hat{A}_y(J^{-1}Q)=\Phi_{\bar{z}}\ , \quad \hat{A}_z(J^{-1}Q)=-\Phi_{\bar{y}} \qquad (17)$$

Equation (17) is a BT connecting the symmetry characteristic $\Phi$ of PSDYM with the symmetry characteristic $Q$ of SDYM. Indeed, the integrability condition $(\Phi_{\bar{z}})_{\bar{y}}=(\Phi_{\bar{y}})_{\bar{z}}$ yields the symmetry condition (13) for SDYM. So, when $Q$ is an SDYM symmetry characteristic, the BT (17) is integrable for $\Phi$. Conversely, the integrability condition $[\hat{A}_z\,,\hat{A}_y](J^{-1}Q)=0$ , valid in view of the zero-curvature condition, yields the PSDYM symmetry condition (14) for $\Phi$ and guarantees integrability for $Q$.

We note that, for a given $Q$, the solution of the BT (17) for $\Phi$ is not unique, and vice versa. To achieve uniqueness we thus need to make some additional assumptions: (*a*) If $\Phi$ is a solution for a given $Q$, then so is $\Phi+M(y,z)$, where $M$ is an arbitrary matrix function. We make the agreement that any arbitrary additive term of the form $M(y,z)$ will be ignored when it appears in the solution for $\Phi$. (*b*) If $Q$ is a solution for a given $\Phi$, then so is $Q+\varepsilon(\bar{y},\bar{z})J$ , where $\varepsilon(\bar{y},\bar{z})$ is an arbitrary matrix function. We agree that any arbitrary additive term of the form $\varepsilon(\bar{y},\bar{z})J$ will be ignored when it appears in the solution for $Q$.

With the above conventions, the BT (17) establishes a 1-1 correspondence between the symmetries of SDYM and those of PSDYM. In particular, the SDYM characteristic $Q$=0 corresponds to the PSDYM characteristic $\Phi$=0. It will be shown below that this correspondence between the two symmetry sets is a Lie algebra isomorphism.

## III. Recursion Operators and Lie-Algebra Isomorphism

Since the two PDEs in (17) are consistent with each other and solvable for $\Phi$ when $Q$ is an SDYM symmetry characteristic, we may use, say, the first equation to formally express $\Phi$ in terms of $Q$:

$$\Phi=D_{\bar{z}}^{-1}\hat{A}_y(J^{-1}Q)\equiv\hat{R}(J^{-1}Q) \qquad (18)$$

where we have introduced the linear operator

$$\hat{R}=D_{\bar{z}}^{-1}\hat{A}_y \qquad (19)$$

***Proposition 1:*** The operator (19) is a recursion operator for PSDYM.

***Proof :*** Let $\Phi$ be a symmetry characteristic for PSDYM. Then, $\Phi$ satisfies the symmetry conditions (14) or (16). We will show that $\Phi'\equiv\hat{R}\Phi$ also is a symmetry characteristic. Indeed,

$$
\begin{aligned}
(\hat{A}_y D_{\bar{y}} + \hat{A}_z D_{\bar{z}})\Phi' &\equiv (\hat{A}_y D_{\bar{y}} + \hat{A}_z D_{\bar{z}})\hat{R}\Phi \\
&= \hat{A}_y D_{\bar{z}}^{-1} D_{\bar{y}} \hat{A}_y \Phi + \hat{A}_z \hat{A}_y \Phi \\
&= \hat{A}_y D_{\bar{z}}^{-1} (D_{\bar{y}} \hat{A}_y + D_{\bar{z}} \hat{A}_z)\Phi + [\hat{A}_z , \hat{A}_y]\Phi = 0 ,
\end{aligned}
$$

in view of (16) and the zero-curvature condition $[\hat{A}_y , \hat{A}_z]=0$. ■

For $sl(N,C)$ PSDYM solutions, the symmetry characteristic $\Phi$ must be traceless. Then, so is the characteristic $\Phi' = \hat{R}\Phi$. That is, the recursion operator (19) preserves the $sl(N,C)$ character of PSDYM.

Is there a systematic process by which one could *derive* the recursion operator (19)? To this end, we seek an auto-BT relating solutions of the PSDYM symmetry condition (14). As shown in [5], such a BT is

$$\hat{A}_y \Phi = \Phi'_{\bar{z}} \quad , \quad \hat{A}_z \Phi = -\Phi'_{\bar{y}} \tag{19a}$$

The first of these equations can then be re-expressed as $\Phi' = \hat{R}\Phi$, with $\hat{R}$ given by (19).

Consider now a symmetry characteristic $Q$ of SDYM, i.e., a solution of the symmetry condition (13). Also, consider the transformation

$$Q' = J\hat{R}\,(J^{-1}Q) \equiv \hat{T}Q \tag{20}$$

where we have introduced the linear operator

$$\hat{T} = J\hat{R}J^{-1} \tag{21}$$

***Proposition 2:*** The operator (21) is a recursion operator for SDYM.

***Proof :*** By assumption, $Q$ is an SDYM symmetry characteristic. Then, as shown above, $\Phi = J^{-1}Q$ is a PSDYM characteristic. Since $\hat{R}$ is a PSDYM recursion operator, $\Phi' \equiv \hat{R}\Phi = \hat{R}(J^{-1}Q)$ also is a PSDYM characteristic. Then, finally, $Q' = J\Phi'$, given by (20), is an SDYM characteristic. ■

For $SL(N,C)$ SDYM solutions, the symmetry characteristic $Q$ must satisfy the condition $tr(J^{-1}Q) = 0$. As can be seen, this condition is preserved by the recursion operator (21). [Note, in this connection, that the BT (17) or (18) properly associates $SL(N,C)$ SDYM characteristics $Q$ with $sl(N,C)$ PSDYM characteristics $\Phi$.]

The recursion operator (21) also can be derived from an auto-BT for the SDYM symmetry condition (13). This BT was constructed in [6] by using a properly chosen Lax pair for SDYM (we refer the reader to this paper for details). We may thus conclude that recursion operators such as (19) or (21) in effect represent auto-BTs for symmetry conditions of respective nonlinear PDEs (see also [7]).

***Lemma:*** The Fréchet derivative $\mathit{\Delta}$ with respect to the vector $V$ of (4), and the recursion operator $\hat{R}$ of (19), satisfy the commutation relation

$$[\mathit{\Delta}, \hat{R}] = D_{\bar{z}}^{-1}\,[\mathit{\Phi}_{\bar{z}}\,,\ ] \tag{22}$$

where $\mathit{\Phi}=\mathit{\Delta}X$, according to (7).

***Proof:*** Introducing an auxiliary function $F$, and using the derivation property (6) of $\mathit{\Delta}$ and the commutativity of $\mathit{\Delta}$ with all total derivatives (as well as all powers of such derivatives), we have:

$$\begin{aligned}\mathit{\Delta}\hat{R}F &= \mathit{\Delta} D_{\bar{z}}^{-1}\hat{A}_y F = D_{\bar{z}}^{-1}\mathit{\Delta}(D_y F + [X_{\bar{z}}\,, F])\\ &= D_{\bar{z}}^{-1}(D_y \mathit{\Delta}F + [(\mathit{\Delta}X)_{\bar{z}}\,, F] + [X_{\bar{z}}\,, \mathit{\Delta}F])\\ &= D_{\bar{z}}^{-1}(\hat{A}_y \mathit{\Delta}F + [\mathit{\Phi}_{\bar{z}}\,, F]) = \hat{R}\,\mathit{\Delta}F + D_{\bar{z}}^{-1}[\mathit{\Phi}_{\bar{z}}\,, F]\,,\end{aligned}$$

from which there follows (22). ■

***Proposition 3:*** The BT (17), or equivalently, its solution (18), establishes an isomorphism between the symmetry Lie algebras of SDYM and PSDYM.

***Proof:*** Let $V$ be a vector field of the form (4), generating a symmetry of the BT (2). As explained previously, since this BT is invariant under $V$, the same will be true with regard to its integrability conditions. Hence, $V$ also represents a symmetry of the SDYM-PSDYM system of equations (1) and (3). The SDYM and PSDYM characteristics are $Q=\mathit{\Delta}J$ and $\mathit{\Phi}=\mathit{\Delta}X$, respectively, where $\mathit{\Delta}$ denotes the Fréchet derivative with respect to $V$. Consider the linear map $I$ defined by (18):

$$I:\ \ \mathit{\Phi} = I\{Q\} = \hat{R}\,J^{-1}Q \tag{23}$$

or

$$I:\ \ \mathit{\Delta}X = I\{\mathit{\Delta}J\} = \hat{R}\,J^{-1}\mathit{\Delta}J \tag{24}$$

Consider also a pair of symmetries of system (2), indexed by $i$ and $j$. These are generated by vector fields $V^{(r)}$, where $r=i,j$. The Fréchet derivatives with respect to the $V^{(r)}$ will be denoted $\mathit{\Delta}^{(r)}$. The SDYM and PSDYM symmetry characteristics are $Q^{(r)}=\mathit{\Delta}^{(r)}J$ and $\mathit{\Phi}^{(r)}=\mathit{\Delta}^{(r)}X$, respectively. According to (24),

$$\mathit{\Delta}^{(r)}X = I\{\mathit{\Delta}^{(r)}J\} = \hat{R}\,J^{-1}\mathit{\Delta}^{(r)}J = \hat{R}\,J^{-1}Q^{(r)}\ ;\quad r=i,j \tag{25}$$

By the Lie-algebraic property of symmetries of PDEs, the functions $[\mathit{\Delta}^{(i)}, \mathit{\Delta}^{(j)}]J$ and $[\mathit{\Delta}^{(i)}, \mathit{\Delta}^{(j)}]X$ also represent symmetry characteristics for SDYM and PSDYM, respectively, where we have put

$$\begin{aligned}[\mathit{\Delta}^{(i)}, \mathit{\Delta}^{(j)}]J &\equiv \mathit{\Delta}^{(i)}\mathit{\Delta}^{(j)}J - \mathit{\Delta}^{(j)}\mathit{\Delta}^{(i)}J = \mathit{\Delta}^{(i)}Q^{(j)} - \mathit{\Delta}^{(j)}Q^{(i)}\,,\\ [\mathit{\Delta}^{(i)}, \mathit{\Delta}^{(j)}]X &\equiv \mathit{\Delta}^{(i)}\mathit{\Delta}^{(j)}X - \mathit{\Delta}^{(j)}\mathit{\Delta}^{(i)}X = \mathit{\Delta}^{(i)}\mathit{\Phi}^{(j)} - \mathit{\Delta}^{(j)}\mathit{\Phi}^{(i)}\,.\end{aligned}$$

We must now verify that

$$[\varDelta^{(i)},\, \varDelta^{(j)}]\, X = I\,\{[\varDelta^{(i)},\, \varDelta^{(j)}]\, J\} = \hat{R}\, J^{-1}[\varDelta^{(i)},\, \varDelta^{(j)}]\, J \qquad (26)$$

Putting $r{=}j$ into (25), and applying the Fréchet derivative $\varDelta^{(i)}$, we have:

$$\begin{aligned} \varDelta^{(i)}\varDelta^{(j)}X &= \varDelta^{(i)}\hat{R}\, J^{-1}Q^{(j)} = [\varDelta^{(i)}, \hat{R}]\, J^{-1}Q^{(j)} + \hat{R}\,\varDelta^{(i)}J^{-1}Q^{(j)} \\ &= D_{\bar{z}}^{-1}[\varPhi_{\bar{z}}^{(i)},\, J^{-1}Q^{(j)}] + \hat{R}\,\varDelta^{(i)}J^{-1}Q^{(j)}, \end{aligned}$$

where we have used the commutation relation (22). By (23) and (19),

$$\varPhi_{\bar{z}}^{(i)} = D_{\bar{z}}\, \hat{R}\, J^{-1}Q^{(i)} = \hat{A}_y\, J^{-1}Q^{(i)}.$$

Moreover, by properties (5) and (8) of the Fréchet derivative,

$$\begin{aligned} \varDelta^{(i)}J^{-1}Q^{(j)} &= -J^{-1}(\varDelta^{(i)}J)\, J^{-1}Q^{(j)} + J^{-1}\varDelta^{(i)}Q^{(j)} \\ &= -\, J^{-1}Q^{(i)}J^{-1}Q^{(j)} + J^{-1}\varDelta^{(i)}Q^{(j)}. \end{aligned}$$

So,

$$\varDelta^{(i)}\varDelta^{(j)}X = D_{\bar{z}}^{-1}[\hat{A}_y J^{-1}Q^{(i)},\, J^{-1}Q^{(j)}] - \hat{R}\, J^{-1}Q^{(i)}J^{-1}Q^{(j)} + \hat{R}\, J^{-1}\varDelta^{(i)}Q^{(j)}.$$

Subtracting from this the analogous expression for $\varDelta^{(j)}\varDelta^{(i)}X$, we have:

$$\begin{aligned} [\varDelta^{(i)},\, \varDelta^{(j)}]X &\equiv \varDelta^{(i)}\varDelta^{(j)}X - \varDelta^{(j)}\varDelta^{(i)}X \\ &= D_{\bar{z}}^{-1}\Big([\hat{A}_y J^{-1}Q^{(i)},\, J^{-1}Q^{(j)}] + [J^{-1}Q^{(i)},\, \hat{A}_y J^{-1}Q^{(j)}]\Big) \\ &\qquad - \hat{R}\,[J^{-1}Q^{(i)},\, J^{-1}Q^{(j)}] + \hat{R}\, J^{-1}(\varDelta^{(i)}Q^{(j)} - \varDelta^{(j)}Q^{(i)}) \\ &= D_{\bar{z}}^{-1}\hat{A}_y\,[J^{-1}Q^{(i)},\, J^{-1}Q^{(j)}] - \hat{R}\,[J^{-1}Q^{(i)},\, J^{-1}Q^{(j)}] \\ &\qquad + \hat{R}\, J^{-1}(\varDelta^{(i)}\varDelta^{(j)}J - \varDelta^{(j)}\varDelta^{(i)}J) \\ &= \hat{R}\, J^{-1}[\varDelta^{(i)},\, \varDelta^{(j)}]\, J \end{aligned}$$

where we have used the derivation property (10) of $\hat{A}_y$ and we have taken (19) into account. Thus, (26) has been proven. ■

Now, suppose $\hat{P}$ is a recursion operator for SDYM, while $\hat{S}$ is a recursion operator for PSDYM. Thus, if $Q$ and $\varPhi$ are symmetry characteristics for SDYM and PSDYM, respectively, then $Q' = \hat{P}Q$ and $\varPhi' = \hat{S}\varPhi$ also are symmetry characteristics.

***Definition:*** The linear operators $\hat{P}$ and $\hat{S}$ will be called *equivalent with respect to the isomorphism I* (or *I-equivalent*) if the following condition is satisfied:

$$\hat{S}\varPhi = I\,\{\hat{P}Q\} \quad when \quad \varPhi = I\,\{Q\} \qquad (27)$$

By using (23), the above condition is written

$$\hat{S}\Phi = \hat{R}J^{-1}\hat{P}Q \quad when \quad \Phi = \hat{R}J^{-1}Q \;\Rightarrow\; \hat{S}\hat{R}J^{-1}Q = \hat{R}J^{-1}\hat{P}Q \ .$$

Thus, in order that $\hat{P}$ and $\hat{S}$ be *I*-equivalent recursion operators, the following operator equation must be satisfied on the infinite-dimensional linear space of all SDYM symmetry characteristics:

$$\hat{S}\hat{R}J^{-1} = \hat{R}J^{-1}\hat{P} \tag{28}$$

Having already found a PSDYM recursion operator $\hat{S}=\hat{R}$, we now want to evaluate the *I*-equivalent SDYM recursion operator $\hat{P}$. To this end, we put $\hat{S}=\hat{R}$ in (28) and write

$$\hat{R}\,(\hat{R}J^{-1} - J^{-1}\hat{P}) = 0 \ .$$

As is easy to see, this is satisfied for $\hat{P} = \hat{T}$, in view of (21). We thus conclude that

- the recursion operators $\hat{R}$ and $\hat{T}$, defined by (19) and (21), are *I*-equivalent.

We note that (28) is a sort of BT relating recursion operators of different PDEs, rather than solutions or symmetries of these PDEs. Thus, if a recursion operator is known for either PDE, this BT will yield a corresponding operator for the other PDE. Note that we have encountered BTs at various levels: (*a*) The non-auto-BT (2), relating solutions of two different nonlinear PDEs (1) and (3); (*b*) the BT (17), or equivalently (18), relating symmetry characteristics of these PDEs; (*c*) the recursion operators (19) and (21), which can be re-expressed as auto-BTs for the symmetry conditions (14) and (13), respectively; and (*d*) the BT (28), relating recursion operators for the original, nonlinear PDEs. (We make the technical observation that the first three BTs are “strong”, while the last one is “weak”; see Appendix.)

***Example:*** Consider the PSDYM symmetry characteristic $\Phi = X_z$ (*z*-translation). To find the *I*-related SDYM characteristic $Q$, we use (23):

$$\hat{R}J^{-1}Q = \Phi \;\Rightarrow\; D_{\bar{z}}^{-1}\hat{A}_y(J^{-1}Q) = X_z \;\Rightarrow\; \hat{A}_y(J^{-1}Q) = X_{z\bar{z}} \;\overset{(2)}{\Rightarrow}$$
$$(J^{-1}Q)_y + [J^{-1}J_y\,,\,J^{-1}Q] = (J^{-1}J_y)_z \ ,$$

which is satisfied for $Q = J_z$. By applying the recursion operator $\hat{T}$ on $Q$,

$$Q' = \hat{T}Q = J\hat{R}J^{-1}Q = JD_{\bar{z}}^{-1}\hat{A}_y(J^{-1}J_z) = JD_{\bar{z}}^{-1}\left\{(J^{-1}J_z)_y + [J^{-1}J_y\,,J^{-1}J_z]\right\}$$
$$= JD_{\bar{z}}^{-1}(J^{-1}J_y)_z \overset{(2)}{=} JD_{\bar{z}}^{-1}X_{z\bar{z}} = JX_z \ .$$

To find the *I*-related PSDYM characteristic $\Phi'$, we use (23) once more:

$$\Phi' = \hat{R}J^{-1}Q' = \hat{R}X_z = \hat{R}\Phi \ .$$

We notice that $\hat{R}\Phi = I\{\hat{T}Q\}$ when $\Phi = I\{Q\}$, as expected by the fact that $\hat{R}$ and $\hat{T}$ are *I*-equivalent recursion operators. ■

Now, let $Q^{(0)}$ be some SDYM symmetry characteristic. By repeated application of the recursion operator $\hat{T}$, we obtain an infinite sequence of such characteristics:

$$Q^{(1)} = \hat{T}Q^{(0)},\ \ Q^{(2)} = \hat{T}Q^{(1)} = \hat{T}^2 Q^{(0)},\ \cdots\ ,\ \ Q^{(n)} = \hat{T}Q^{(n-1)} = \hat{T}^n Q^{(0)},\ \cdots$$

(we note that any power of a recursion operator also is a recursion operator). Also, let

$$\Phi^{(0)} = I\{Q^{(0)}\} = \hat{R}J^{-1}Q^{(0)} \qquad (29)$$

be the PSDYM characteristic which is *I*-related to $Q^{(0)}$. Repeated application of the PSDYM recursion operator $\hat{R}$ will now yield an infinite sequence of PSDYM characteristics. Taking into account that $\hat{R}$ and $\hat{T}$ are *I*-equivalent recursion operators, we can write this sequence as follows:

$$\Phi^{(1)} = \hat{R}\Phi^{(0)} = I\{\hat{T}Q^{(0)}\}\ ,\quad \Phi^{(2)} = \hat{R}^2\Phi^{(0)} = I\{\hat{T}^2 Q^{(0)}\}\ ,\ \cdots\ ,$$
$$\Phi^{(n)} = \hat{R}^n\Phi^{(0)} = I\{\hat{T}^n Q^{(0)}\}\ ,\ \cdots$$

Assume now that the infinite set of SDYM symmetries represented by the characteristics $\{Q^{(n)}\}$ $(n = 0,1,2,\cdots)$ has the structure of a Lie algebra. This set then constitutes a symmetry subalgebra of SDYM. Given that the set $\{\Phi^{(n)}\}$ is *I*-related to $\{Q^{(n)}\}$ and that *I* is a Lie-algebra isomorphism, we conclude that the infinite set of characteristics $\{\Phi^{(n)}\}$ corresponds to a symmetry subalgebra of PSDYM which is isomorphic to the associated subalgebra $\{Q^{(n)}\}$ of SDYM.

More generally, let $\{Q_k^{(0)} \,/\, k = 1,2,\cdots,p\}$ be a finite set of SDYM symmetry characteristics, and let $\{\Phi_k^{(0)} \,/\, k = 1,2,\cdots,p\}$ be the *I*-related set of PSDYM characteristics, where

$$\Phi_k^{(0)} = I\{Q_k^{(0)}\} = \hat{R}J^{-1}Q_k^{(0)}\ ;\ \ k = 1,2,\cdots,p \qquad (30)$$

Assume that the infinite set of characteristics

$$\{Q_k^{(n)} = \hat{T}^n Q_k^{(0)} \,/\, n = 0,1,2,\cdots\ ;\ k = 1,2,\cdots,p\} \qquad (31)$$

corresponds to a Lie subalgebra of SDYM symmetries. Then, the *I*-related set of characteristics

$$\{\Phi_k^{(n)} = \hat{R}^n \Phi_k^{(0)} \,/\, n = 0,1,2,\cdots\ ;\ k = 1,2,\cdots,p\} \qquad (32)$$

corresponds to a PSDYM symmetry subalgebra which is isomorphic to that of (31).

Let us summarize our main conclusions:

- The infinite-dimensional symmetry Lie algebras of SDYM and PSDYM are isomorphic, the isomorphism $I$ being defined by (23) or (24).
- The recursion operators $\hat{T}$ and $\hat{R}$, defined in (21) and (19), when applied to $I$-related symmetry characteristics [such as those in (29) or (30)], may generate isomorphic, infinite-dimensional symmetry subalgebras of SDYM and PSDYM, respectively.
- Since the structures of the symmetry Lie algebras of SDYM and PSDYM are similar, all results regarding the latter structure are also applicable to the SDYM case.

***Comment:*** At this point the reader may wonder whether it is really necessary to go through the PSDYM symmetry problem in order to solve the corresponding SDYM problem. In principle, of course, the SDYM case can be treated on its own. In practice, however, it is easier to study the symmetry structure of PSDYM first and then take advantage of the isomorphism between this structure and that of SDYM. This statement is justified by the fact that the PSDYM recursion operator is considerably easier to handle compared to the corresponding SDYM operator. This property of the former operator is of great value in the interest of computational simplicity (in particular, for the purpose of deriving various commutation relations; cf. [8]).

## IV. Potential Symmetries and Current Algebras

We recall that every SDYM symmetry characteristic can be expressed as $Q=J\Phi$, where $\Phi$ is a PSDYM characteristic (we note that $\Phi$ is *not* $I$-related to $Q$). Let $\Phi$ be a characteristic which depends locally or nonlocally on $X$ and/or various derivatives of $X$. By the BT (2), $X$ must be an integral of $J$ and its derivatives, and so it and its derivatives $X_y$ and $X_z$ are nonlocal in $J$. On the other hand, according to (2), the quantities $X_{\bar{y}}$ and $X_{\bar{z}}$ depend locally on $J$. Thus, in general, $\Phi$ can be local or nonlocal in $J$. In the case where $\Phi$ is *nonlocal* in $J$, we say that the characteristic $Q=J\Phi$ expresses a *potential symmetry* of SDYM [3, 5]. (See Appendix for a general definition of locality and nonlocality of symmetries.)

***A. Internal Symmetries.*** The PSDYM equation is generally invariant under a transformation of the form

$$\Delta^{(0)}X = \Phi^{(0)} = [X, M] \qquad (33)$$

where $M$ is any constant $sl(N,C)$ matrix. Since the characteristic $\Phi^{(0)}$ is nonlocal in $J$, the transformation

$$Q = J\Phi^{(0)} = J\,[X, M]$$

is a genuine potential symmetry of SDYM. Note that the SDYM characteristic which is $I$-related to $\Phi^{(0)}$ is not $Q$, but rather $Q^{(0)} = JM$, since we then have

$$\hat{R}J^{-1}Q^{(0)} = \hat{R}M = D_{\bar{z}}^{-1}[X_{\bar{z}}, M] = [X, M] = \Phi^{(0)} .$$

Let $\{\tau_k\}$ be a basis for $sl(N,C)$:

$$[\tau_i, \tau_j] = C_{ij}^k \tau_k .$$

Then $M$ is expanded as $M = \alpha^k \tau_k$, and (33) is resolved into a set of independent basis transformations

$$\Delta_k^{(0)} X = \Phi_k^{(0)} = [X, \tau_k]$$

corresponding to the SDYM potential symmetries

$$Q_k = J\Phi_k^{(0)} = J[X, \tau_k] .$$

These are not the same as the $I$-related characteristics

$$\Delta_k^{(0)} J = Q_k^{(0)} = J\tau_k .$$

Consider now the infinite set of transformations

$$\Delta_k^{(n)} X = \Phi_k^{(n)} = \hat{R}^n \Phi_k^{(0)} = \hat{R}^n [X, \tau_k] \quad (n = 0, 1, 2, \cdots) \qquad (34)$$

As can be shown, they satisfy the commutation relations of a Kac-Moody algebra:

$$[\Delta_i^{(m)}, \Delta_j^{(n)}] X = C_{ij}^k \Delta_k^{(m+n)} X .$$

In view of the isomorphism $I$, this structure is also present in SDYM. Indeed, this is precisely the familiar hidden symmetry of SDYM [11, 12]. The SDYM transformations which are $I$-related to those in (34) are given by

$$\Delta_k^{(n)} J = Q_k^{(n)} = \hat{T}^n Q_k^{(0)} = \hat{T}^n J\tau_k \quad (n = 0, 1, 2, \cdots) .$$

They constitute an infinite set of potential symmetries (note, for example, that $\Delta_k^{(1)} J = J[X, \tau_k] = J\Phi_k^{(0)}$) and they satisfy the commutation relations

$$[\Delta_i^{(m)}, \Delta_j^{(n)}] J = C_{ij}^k \Delta_k^{(m+n)} J .$$

***B. Symmetries in the Base Space.*** A number of local PSDYM symmetries corresponding to coordinate transformations are nonlocal in $J$, hence lead to potential symmetries of SDYM. By using isovector methods [4, 15], nine such PSDYM symmetries can be found. They can be expressed as follows:

$$\varDelta_k^{(0)} X = \varPhi_k^{(0)} = \hat{L}_k X \quad (k = 1, 2, \cdots, 9) \tag{35}$$

where the $\hat{L}_k$ are nine linear operators which are explicitly given by

$$\hat{L}_1 = D_y \ , \quad \hat{L}_2 = D_z \ , \quad \hat{L}_3 = zD_y - \overline{y} D_{\overline{z}} \ , \quad \hat{L}_4 = yD_z - \overline{z} D_{\overline{y}} \ ,$$
$$\hat{L}_5 = yD_y - zD_z - \overline{y} D_{\overline{y}} + \overline{z} D_{\overline{z}} \ , \quad \hat{L}_6 = 1 + yD_y + zD_z \ ,$$
$$\hat{L}_7 = 1 - \overline{y} D_{\overline{y}} - \overline{z} D_{\overline{z}} \ , \quad \hat{L}_8 = y\hat{L}_6 + \overline{z}\,(yD_{\overline{z}} - zD_{\overline{y}}) \ ,$$
$$\hat{L}_9 = z\hat{L}_6 + \overline{y}\,(zD_{\overline{y}} - yD_{\overline{z}}) \ .$$

The $\hat{L}_1$, $\hat{L}_2$ represent translations of $y$ and $z$, respectively, while the $\hat{L}_3$, $\hat{L}_4$ represent rotational symmetries. The $\hat{L}_5$, $\hat{L}_6$, $\hat{L}_7$ express scale transformations, while $\hat{L}_8$ and $\hat{L}_9$ represent nonlinear coordinate transformations which presumably reflect the special conformal invariance of the SDYM equations in their original, covariant form.

The first five operators $\hat{L}_1, \cdots, \hat{L}_5$ form the basis of a Lie algebra, the commutation relations of which we write in the form

$$[\hat{L}_i \, , \, \hat{L}_j] = -\, f_{ij}^{\,k} \, \hat{L}_k \quad (k = 1, \cdots, 5) \ .$$

Consider now the infinite set of transformations

$$\varDelta_k^{(n)} X = \varPhi_k^{(n)} = \hat{R}^n \varPhi_k^{(0)} = \hat{R}^n \hat{L}_k X \quad (k = 1, \cdots, 5) \tag{36}$$

As can be shown [8], these form a Kac-Moody algebra:

$$[\varDelta_i^{(m)}, \, \varDelta_j^{(n)}] X = f_{ij}^{\,k} \, \varDelta_k^{(m+n)} X \ .$$

Consider also the infinite sets of transformations

$$\varDelta^{(n)} X = \hat{R}^n \hat{L}_6 X \quad \text{and} \quad \varDelta^{(n)} X = \hat{R}^n \hat{L}_7 X \tag{37}$$

As can be proven [8], each set forms a Virasoro algebra (apart from a sign):

$$[\varDelta^{(m)}, \, \varDelta^{(n)}] X = -\,(m - n)\, \varDelta^{(m+n)} X \ .$$

Taking the isomorphism $I$ into account, we conclude that the SDYM symmetry algebra possesses both Kac-Moody and Virasoro subalgebras ("current algebras" [16]), both of which are associated with infinite sets of potential symmetries. The former subalgebras are associated with both internal and coordinate transformations, while the latter ones are related to coordinate transformations only. These conclusions are in agreement with those of [13], although the mathematical approach there is different from ours.

## V. Summary

By using the SDYM-PSDYM system as a model, we have studied a process for associating symmetries and recursion operators of two nonlinear PDEs related to each other by a non-auto-BT. The concept of a BT itself enters our analysis at various levels: (*a*) The non-auto BT (2) relates solutions of the nonlinear PDEs (1) and (3); (*b*) the non-auto-BT (17) or (18) relates symmetry characteristics of these PDEs; (*c*) the auto-BTs for the symmetry conditions (14) and (13) lead to the recursion operators (19) and (21), respectively; and (*d*) the transformation (28) may be perceived as a BT associating recursion operators for the original, nonlinear PDEs. We have proven the isomorphism between the infinite-dimensional symmetry Lie algebras of SDYM and PSDYM, and we have used this property to draw several conclusions regarding the Lie-algebraic structure of the potential symmetries of SDYM.

For further reading on recursion operators, the reader is referred to [17-22]. A nice discussion of the SDYM symmetry structure and its connection to the existence of infinitely many conservation laws can be found in the paper by Adam *et al.* [23].

## VI. Appendix: Some Basic Definitions

To make the paper as self-contained as possible, basic definitions of some key concepts that are being used are given below:

***A. Recursion Operators.*** Consider a PDE $F[u]=0$, in the dependent variable $u$ and the independent variables $x^\mu$ ($\mu=1,2,\ldots$). Let $\delta u=\alpha Q[u]$ be an infinitesimal symmetry transformation of the PDE, where $Q[u]$ is the *symmetry characteristic*. The symmetry is generated by the (formal) vector field

$$V=Q[u]\frac{\partial}{\partial u}+\ prolongation\ =Q\frac{\partial}{\partial u}+Q_\mu\frac{\partial}{\partial u_\mu}+Q_{\mu\nu}\frac{\partial}{\partial u_{\mu\nu}}+\cdots \qquad \text{(A.1)}$$

(where the $Q_\mu\equiv D_\mu Q$, etc., denote total derivatives of $Q$). The symmetry condition is expressed by a PDE, linear in $Q$:

$$S(Q;u)\equiv \Delta F[u]=0 \ \bmod\ F[u] \qquad \text{(A.2)}$$

where $\Delta$ denotes the Fréchet derivative with respect to $V$. If $u$ is a scalar quantity, then (A.2) takes the form

$$S(Q;u)=VF[u]=Q\frac{\partial F}{\partial u}+Q_\mu\frac{\partial F}{\partial u_\mu}+Q_{\mu\nu}\frac{\partial F}{\partial u_{\mu\nu}}+\cdots=0 \ \bmod\ F[u] \qquad \text{(A.3)}$$

Since the PDE (A.2) is linear in $Q$, the sum of two solutions (for the same $u$) also is a solution. Thus, for any given $u$, the solutions $\{Q[u]\}$ of (A.2) form a linear space $S_u$. A *recursion operator* $\hat{R}$ is a linear operator which maps the space $S_u$ into itself. Thus, if $Q$ is a symmetry characteristic of $F[u]=0$ [i.e., a solution of (A.2)], then so is $\hat{R}Q$:

$$S(\hat{R}Q;u)=0 \quad when \quad S(Q;u)=0 \qquad \text{(A.4)}$$

We note that $\hat{R}^2Q,\hat{R}^3Q,\cdots,\hat{R}^nQ$ also are symmetry characteristics. This means that

- any power $\hat{R}^n$ of a recursion operator also is a recursion operator.

Thus, starting with any symmetry characteristic $Q$, we can obtain an infinite set of such characteristics by repeated application of the recursion operator.

A *symmetry operator* $\hat{L}$ is a linear operator, independent of $u$, which produces a symmetry characteristic $Q[u]$ when it acts on $u$. Thus, $\hat{L}u = Q[u]$. We note that $\hat{R}\hat{L}u$ is a symmetry characteristic, which means that

- the product $\hat{R}\hat{L}$ of a recursion operator and a symmetry operator is a symmetry operator.

Thus, given that $\hat{R}^n$ is a recursion operator, we conclude that $\hat{R}^n\hat{L}u$ is a member of $S_u$. Examples of symmetry operators are the nine operators $\hat{L}_k$ that appear in (35), as well as the operator $\hat{L} = [\ \ , M\,]$ which is implicitly defined in (33).

***B. "Strong" and "Weak" Bäcklund Transformations.*** In the most general sense, a BT is a set of relations (typically differential, although in certain cases algebraic ones are also considered) which connect solutions of two different PDEs (non-auto-BT) or of the same PDE (auto-BT). The technical distinction between "strong" and "weak" BTs [24, 25] can be roughly described as follows: In a strong BT connecting, say, the variables $u$ and $v$, integrability of the differential system for either variable *demands* that the other variable satisfy a certain PDE. A weak BT, on the other hand, is much like a symmetry transformation: $u$ and $v$ are not, *a priori*, required to satisfy any particular PDEs for integrability. *If*, however, $u$ satisfies some specific PDE, *then* $v$ satisfies some related PDE. (An example is the Cole-Hopf transformation, connecting solutions of Burgers' equation to solutions of the heat equation.)

The BT (2) is strong, since its integrability conditions *force* the functions $J$ and $X$ to satisfy the PDEs (1) and (3), respectively. Similar remarks apply to the BTs (17) and (19*a*). On the other hand, transformation (28) does not *a priori* impose any specific properties on the operators $\hat{P}$ and $\hat{S}$. *If*, however, $\hat{P}$ is an SDYM recursion operator, *then* $\hat{S}$ is the *I*-equivalent PSDYM recursion operator. Thus, equation (28) is a Bäcklund-like transformation of the weak type, although this particular transformation relates operators rather than functions.

***C. Local and Nonlocal Symmetries.*** Let $F[u]=0$ be a PDE in the dependent variable $u$ and the independent variables $x^\mu$ ($\mu$=1,2,...). A symmetry characteristic $Q[u]$ represents a *local* symmetry of the PDE if $Q$ depends, at most, on $x^\mu$, $u$, and derivatives of $u$ with respect to the $x^\mu$. A symmetry is *nonlocal* if the corresponding characteristic $Q$ contains additional variables expressed as *integrals* of $u$ with respect to the $x^\mu$ (or, more generally, integrals of local functions of $u$). As an example, the PSDYM characteristic $\Phi = [X, M\,]$ (where $M$ is a constant matrix) represents a local symmetry of this PDE (since it depends locally on the PSDYM variable $X$), whereas the SDYM characteristic $Q = J\,[X, M\,]$ represents a nonlocal symmetry of that PDE since it contains an additional variable $X$ which is expressed as an integral of a local function of the principal SDYM variable $J$ [this follows from the BT (2)]. The infinite symmetries (34), (36) and (37) are increasingly nonlocal in $X$ for $n$>0, since they are produced by repeated application of the integro-differential recursion operator $\hat{R}$.

## Acknowledgment

We appreciate the kind assistance of Kathleen O'Shea-Arapoglou.